\documentclass[aip,pop,reprint,amsmath,amssymb,english]{revtex4-1}

\usepackage{graphicx}
\usepackage[nooneline]{subfigure} 
\usepackage{dcolumn}
\usepackage{bm}
\usepackage{color}
\usepackage{textcomp}
\usepackage{babel}
\usepackage{csquotes}
\usepackage[normalem]{ulem}

\MakeOuterQuote{"}

\begin{document}

\title{Selective Deuterium Ion Acceleration Using the Vulcan PW Laser}

\author{A.G. Krygier}
\affiliation{Laboratoire pour l'Utilisation des Lasers Intenses, \'{E}cole Polytechnique, 91128 Palasiseau, France}
\affiliation{Physics Department, The Ohio State University, Columbus, Ohio, 43210, USA}

\author{J.T. Morrison}
\affiliation{Propulsion Systems Directorate, Air Force Research Lab, Wright Patterson Air Force Base, Ohio 45433, USA}

\author{S. Kar}
\email{s.kar@qub.ac.uk}
\affiliation{Centre for Plasma Physics, School of Mathematics and Physics, Queen’s University Belfast, Belfast BT7 1NN, United Kingdom}

\author{H. Ahmed}
\affiliation{Centre for Plasma Physics, School of Mathematics and Physics, Queen’s University Belfast, Belfast BT7 1NN, United Kingdom}

\author{A. Alejo}
\affiliation{Centre for Plasma Physics, School of Mathematics and Physics, Queen’s University Belfast, Belfast BT7 1NN, United Kingdom}

\author{R. Clarke}
\affiliation{Central Laser Facility, Rutherford Appleton Laboratory, Didcot, Oxfordshire OX11 0QX, United Kingdom}

\author{J. Fuchs}
\affiliation{Laboratoire pour l'Utilisation des Lasers Intenses, \'{E}cole Polytechnique, 91128 Palasiseau, France}

\author{A. Green}
\affiliation{Centre for Plasma Physics, School of Mathematics and Physics, Queen’s University Belfast, Belfast BT7 1NN, United Kingdom}

\author{D. Jung}
\affiliation{Centre for Plasma Physics, School of Mathematics and Physics, Queen’s University Belfast, Belfast BT7 1NN, United Kingdom}

\author{A. Kleinschmidt}
\affiliation{Institut f\"{u}r Kernphysik, Technische Universit\"{a}t Darmstadt, Schlo$\beta$gartenstrasse 9, D-64289 Darmstadt, Germany}

\author{Z. Najmudin}
\affiliation{The John Adams Institute, Blackett Laboratory, Department of Physics, Imperial College London SW7 2AZ, United Kingdom}

\author{H. Nakamura}
\affiliation{The John Adams Institute, Blackett Laboratory, Department of Physics, Imperial College London SW7 2AZ, United Kingdom}

\author{P. Norreys}
\affiliation{Central Laser Facility, Rutherford Appleton Laboratory, Didcot, Oxfordshire OX11 0QX, United Kingdom}
\affiliation{Department of Physics, University of Oxford, Oxford OX1 3PU, United Kingdom}

\author{M. Notley}
\affiliation{Central Laser Facility, Rutherford Appleton Laboratory, Didcot, Oxfordshire OX11 0QX, United Kingdom}

\author{M. Oliver}
\affiliation{Department of Physics, University of Oxford, Oxford OX1 3PU, United Kingdom}

\author{M. Roth}
\affiliation{Institut f\"{u}r Kernphysik, Technische Universit\"{a}t Darmstadt, Schlo$\beta$gartenstrasse 9, D-64289 Darmstadt, Germany}

\author{L. Vassura}
\affiliation{Laboratoire pour l'Utilisation des Lasers Intenses, \'{E}cole Polytechnique, 91128 Palasiseau, France}

\author{M. Zepf}
\affiliation{Centre for Plasma Physics, School of Mathematics and Physics, Queen’s University Belfast, Belfast BT7 1NN, United Kingdom}
\affiliation{Helmholtz Institut Jena, D-07743 Jena, Germany}

\author{M. Borghesi}
\affiliation{Centre for Plasma Physics, School of Mathematics and Physics, Queen’s University Belfast, Belfast BT7 1NN, United Kingdom}
\affiliation{Institute of Physics of the ASCR, ELI-Beamlines Project, Na Slovance 2, 18221 Prague, Czech Republic}

\author{R.R. Freeman}
\affiliation{Physics Department, The Ohio State University, Columbus, Ohio, 43210, USA}

\date{\today}

\begin{abstract}
	
We report on the successful demonstration of selective acceleration of deuterium ions by target-normal sheath acceleration (TNSA) with a high-energy petawatt laser. TNSA typically produces a multi-species ion beam that originates from the intrinsic hydrocarbon and water vapor contaminants on the target surface. Using the method first developed by Morrison, et al., \cite{Morrison:POP2012} an ion beam with $>$99$\%$ deuterium ions and peak energy 14 MeV/nucleon is produced with a 200 J, 700fs, $>10^{20} W/cm^{2}$ laser pulse by cryogenically freezing heavy water (D$_{2}$O) vapor onto the rear surface of the target prior to the shot. Within the range of our detectors (0-8.5$^{\circ}$), we find laser-to-deuterium-ion energy conversion efficiency of 4.3$\%$ above 0.7 MeV/nucleon while a conservative estimate of the total beam gives a conversion efficiency of 9.4$\%$. 

\end{abstract}

\maketitle

\section{Introduction}

The acceleration of ions with high-power lasers has drawn a great deal of attention over the last 15 years. This interest has been driven by a wide range of promising applications \cite{Macchi:RMP2013} coupled with the constantly improving capabilities of laser facilities. Bunches of MeV ions can be used for radiobiological studies possibly relevant for cancer therapy,\cite{Bulanov:PLA2002,Doria:AIP2012} creation of warm dense matter with isochoric heating,\cite{Patel:PRL2003} proton driven fast ignition,\cite{Roth:PRL2001} and neutron production.  The best known and most investigated mechanism for generating fast ion beams with a high-intensity laser is target-normal sheath acceleration (TNSA),\cite{Snavely:PRL2000, Wilks:POP2001} which produces protons with 10's MeV energies. The ions are accelerated by the sheath field that is formed on the rear surface of the target by laser-generated hot electrons. \cite{Wilks:1992,Pukhov:1998,Gahn:1999,Arefiev:2012,Krygier:POP2014,Jiang:PRE2014} TNSA ion beams typically have a 
broad distribution in energy  and emission angle with laser-to-ion conversion efficiency up to several percent.\cite{Robson:Nature2007,Daido:RPP2012}

Of particular recent interest is the generation of neutron beams from laser-irradiated targets; one mechanism for laser-based neutron production is the pitcher-catcher scheme.\cite{Fritzler:PRL2002} Here, the laser produces an ion beam (typically  protons or deuterium ions) from a primary "pitcher" target. The ions are collided into a secondary "catcher" target that is composed of a fusable material. There are several reactions which can produce neutrons in this configuration including p(Li,n), d(d,n), d(t,n), d(Li,n), and d(Be,n). The deuterium cross sections are all larger than p(Li,n) but, without special care, TNSA predominantly accelerates protons; until recently, progress in developing a deuterium source that can utilize the larger cross sections has been limited.

There are several emerging, alternative mechanisms for the production of intense particle beams from laser-plasma interactions: e.g., radiation pressure acceleration (RPA)\cite{Esirkepov:PRL2004, Kar:PRL2012} and breakout afterburner (BOA).\cite{Yin:LPB2006} However, these mechanisms have strict laser and target requirements and are still under active development. Currently, the simplest configuration to produce a deuterium beam for neutron generation is TNSA from a deuterium-rich target (deuterated plastic, for example). However, the ion beams produced by this configuration are ubiquitously dominated by C$^{+}$, O$^{+}$, and H$^{+}$;\cite{Morrison:POP2012,Higginson:POP2011} apparently, controlling the accelerated ion species takes more than target material selection. The reason for this is well known: sub-micron layers of hydrocarbon and water-vapor contaminants 
cover the targets. The difficulty in satisfactorily employing the pitcher-catcher technique for neutron production has been discussed by Willingale, et al. 
\cite{Willingale:POP2011}

There have been several attempts at reducing the yield of contaminant ions in TNSA. Heating a 50 $\mu$m thick Al target to 600 K has been shown to reduce the peak energy and yield of protons by about an order of magnitude while enhancing species coated onto the substrate. \cite{Hegelich:PRL2002,Offerman:POP2011} Ablation with a secondary laser\cite{Mackinnon:PRL2001} has also been shown to reduce the proton signal by decreasing the rear surface field.  Unfortunately, none of these contaminant reduction schemes have produced the desired quality of deuterium ion sources. 

Here we report the results of an experiment that extends the method developed by Morrison, et al.,\cite{Morrison:POP2012} to a new regime of laser energy and intensity.  This approach freezes a $\mu$m's thick layer of heavy water over the ubiquitous proton-rich contaminants. Morrison demonstrated that this method produces an ion beam with $\sim 99\%$ deuterium ions while maintaining typical laser-to-ion conversion efficiency. The ice layer addition can be done quickly and reproducibly, and is synchronized with the laser to prevent regrowth of contaminants.  We have produced an ion beam that has a similar deuterium-to-proton ratio to the heavy water ($>$0.99) and high laser-to-deuterium-ion conversion efficiency.

\section{Experimental Setup}

This experiment was performed using the petawatt arm of the Vulcan laser at the Rutherford Appleton Laboratory (RAL). As configured for this experiment, the laser delivered 200 J on target in a 700 fs pulse of 1053 nm light and is focused to above 10$^{20}$ W/cm$^{2}$ with an $f$/3 off-axis parabola to a $\sim$6 $\mu$m full-width at half-maximum spot.  The laser is normally incident off of a plasma mirror onto the front surface of a 10 $\mu$m thick Au foil with a 3 $\mu$m ($\pm 1$ $\mu$m) thick layer of frozen heavy water on the rear surface.

\subsection{Ice Formation}

\begin{figure}[thb]
	\includegraphics[width=0.42\textwidth,natwidth=1314,natheight=940]{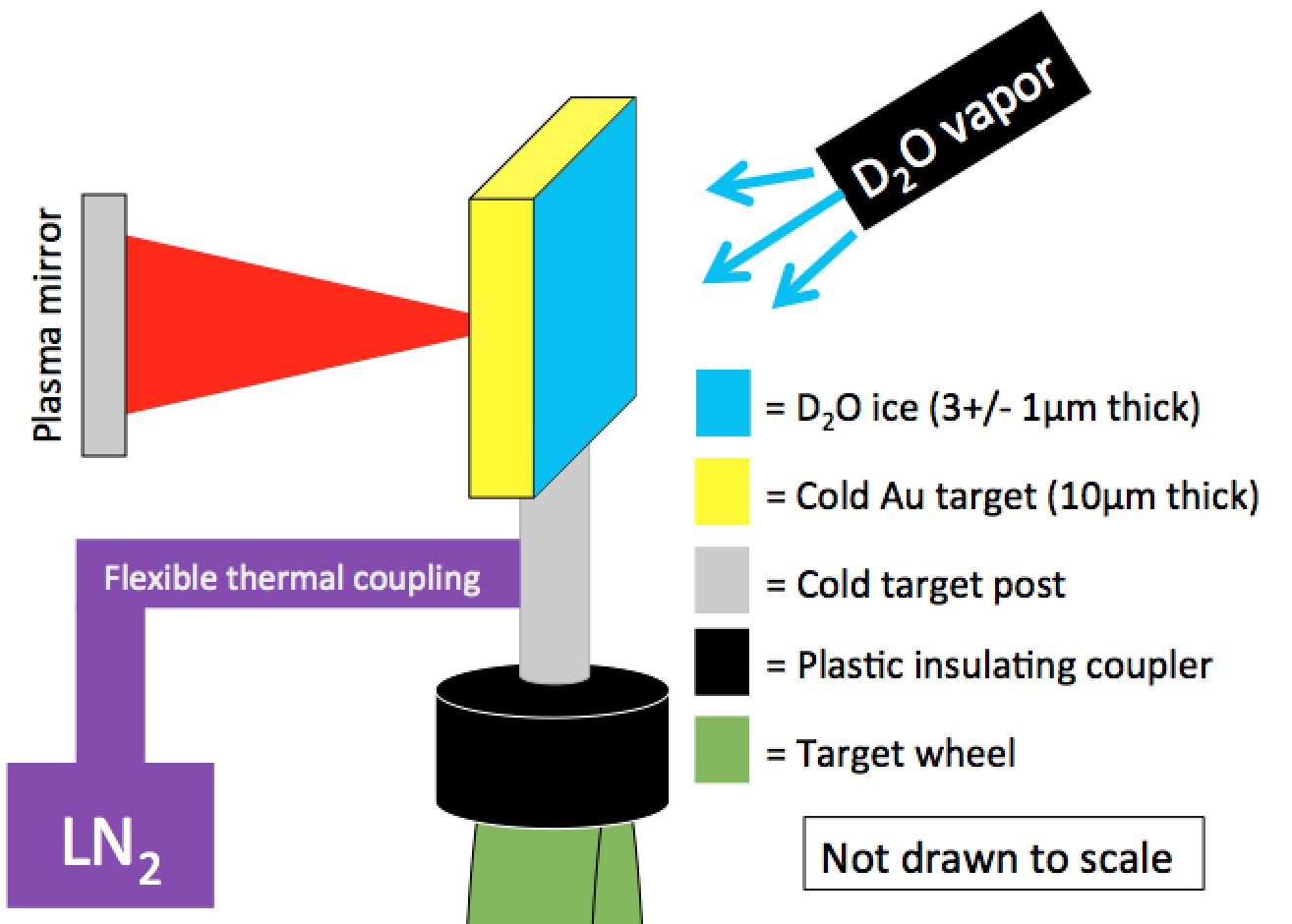}
	\caption{(color online). The target configuration for freezing heavy water on a 10 $\mu$m thick Au foil target. The laser pulse reflects off of the plasma mirror and is normally incident on the target (shown in yellow). The heavy water vapor (blue) comes in through a standard 6 mm outer diameter tube (black) whose outlet is $\sim$6 cm from the Au target. The target is mounted with an Al post that is thermally isolated from the vacuum chamber and target stage (green) by an insulating coupler (black). The target is cooled to below -100 C by a liquid nitrogen reservoir (purple) which is flexibly connected (also purple) to the target post.}
	\label{fig:gas_setup}
\end{figure}

Figure \ref{fig:gas_setup} shows the in-chamber portion of the setup used in this experiment. The ice layer is formed by cryogenically cooling the Au target and releasing a small puff of heavy water vapor just before the laser shot. The vapor source is a trapped volume of heavy water that had a measured ambient vapor pressure of 11 Torr. The outlet nozzle was positioned $\sim$6 cm from the target at a $\sim$50$^{\circ}$ angle with respect to the target surface. The vapor is let into the chamber just before the shot with a solenoidal valve that is synchronized with the the laser system. For the data presented, we used the maximum possible growth rate 
of $\sim 1 \mu$m/s, found empirically to produce the best deuterium beams. $\sim 3\mu$m was the minimum achieveable thickness due to technical reasons related to triggering. Thinner ice layers, which may produce even higher TNSA laser-to-ion conversion efficiency, \cite{Mackinnon:PRL2002,Flacco:PRE2010} could be readily produced with improved timing.

The heavy ice thickness was characterized with dynamic thin-film interference reflectometry. The puff duration-to-ice thickness calibration was done pre-shot - a conflict of diagnostics within the target chamber prevented in situ characterization. A visible cw laser is reflected off of the surface and onto a charge-coupled device. As the ice grows, constructive or destructive interference peaks occur when the optical path is equal to alternating half-integer multiples of the laser wavelength. Counting the number of these peaks that occur during the growth gives the total thickness. Critically, the targets are coated with a surfactant which enables optically smooth ice formation. The surfactant is applied outside of the target chamber before the shot and optically smooth ice can be grown even after several hours in vacuum. Thorough pre-shot testing indicated that the ice layering was repeatable for the same conditions of heavy water vapor pressure in our apparatus and growth duration.

\subsection{Ion Spectra Characterization}

The ion spectra are recorded with 4 Thomson parabola spectrometers (TPS) at different angles (-6$^{\circ}$, 0$^{\circ}$, 3.5$^{\circ}$, and 8.5$^{\circ}$); there was an additional TPS at 30$^{\circ}$ that recorded no ion signal. Each TPS has a 100 $\mu$m radius pinhole which subtend between 1.8-2.1$\times10^{-8}$ sr. The ions are detected with BAS-TR \cite{fujifilm} image plates (IP). A previously ubiquitous problem in this type of experiment is segregating the proton and deuterium ion signals from the higher mass contaminants. There are two causes for this. First, the TPS disperses ions by their charge-to-mass ratio and so, for example, the deuterium ion (D$^{1+}$), C$^{6+}$, and O$^{8+}$, etc. ion signals are overlapped. Second, the highest energies of neighboring charge-to-mass species can be overlapped due to insufficient dispersion and non-zero pinhole diameter. Here, the overlapping of heavier ions with the deuterium and proton signals is prevented by using differential filters for the IPs.

\begin{figure}
	\includegraphics[width=0.42\textwidth,natwidth=656,natheight=672]{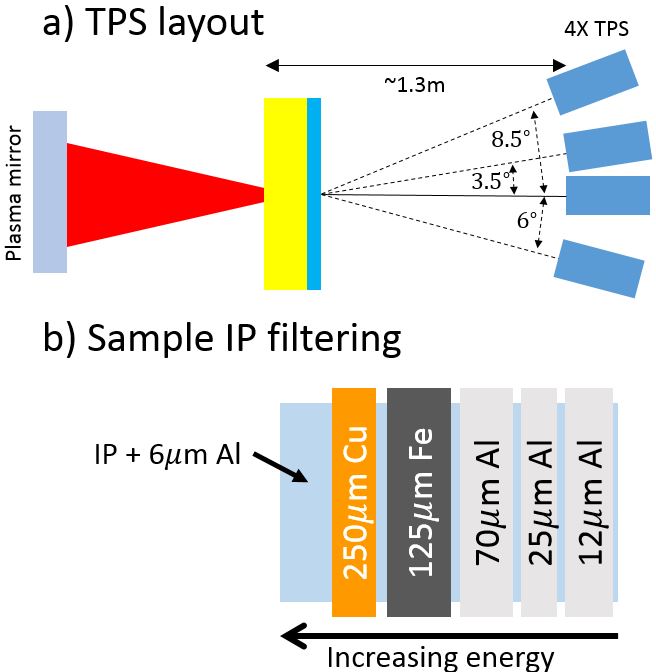}
	\caption{(color online). The detector arrangement is shown in (a); there was an additional TPS at 30$^{\circ}$ that recorded no ion signal. The 100 $\mu$m radius pinholes for the TPS are all located between 1.23-1.31 m from the target and subtend between 1.8-2.1$\times 10^{-8}$ sr. An example of the differential IP filtering is shown in (b). The filters are chosen such that heavier ions like C$^{+}$ and O$^{+}$ are stopped before reaching the IP while light ions (H$^{+}$ and D$^{+}$) are tramsitted with reduced kinetic energy. The entire IP is covered in 6 $\mu$m thick Al so the region labeled 12 $\mu$m Al is filtered with a total of 18 $\mu$m Al, etc.}
	\label{fig:TPS_setup}
\end{figure}

The layout and an example of the differential filtering is shown in Figure \ref{fig:TPS_setup}. The details of the filtering are discussed thoroughly by Alejo, et al. \cite{Alejo:RSI2014} The thicknesses and materials of the IP flters are chosen to stop the higher-mass contaminant ions while allowing the deuterium ions and protons through to the IP. The entire IP is covered with a 6 $\mu$m thick Al foil so the 12 $\mu$m Al filter region is actually filtered by 18 $\mu$m thick Al and so on.

The energy dispersion onto the image plates is calculated numerically using a previously benchmarked method \cite{Morrison:RSI2011} which accounts for fringe fields that are normally not taken into account for this type of diagnostic. The TPS consists of two regions of separate and parallel static magnetic and electric fields which are achieved by a yoked permanent magnet pair (peak 1.02 T) and parallel conducting plates with a large potential difference (16 kV, 15mm separation). We use RADIA, \cite{RADIA:manual} a three dimensional magnetostatic solver, to characterize the magnetic fields - basic features are double checked with a Teslameter. Laplace's equation is solved using Matlab's built in partial differential equation solver which gives the electric field in two dimensions. Test particles are numerically propagated through this system and are spatially mapped onto the IP giving the energy dispersion for each TPS. The absolute response and decay properties of the IP are described in Alejo, et al. \cite{Alejo:RSI2014}.

In some cases, the deuterium signal was so strong that the scanner recorded regions of saturation. The PSL values for these pixels are calculated by rescanning the IP to generate a calibration function. The saturated values are extrapolated using this function; an example of this is in Figure \ref{fig:saturation}, which shows the correction for the scan shown below in Figure \ref{fig:TP3_raw_spec}.

\begin{figure} [h]
\includegraphics[width=0.4\textwidth,natwidth=900,natheight=650]{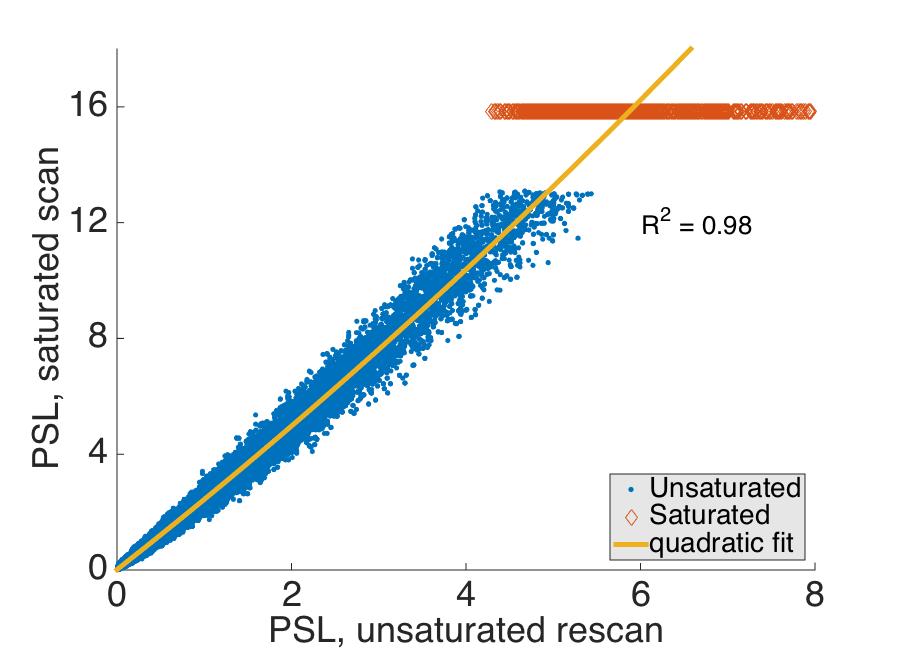}
\caption{(color online). Sample correction curve for a saturated scan: the unsaturated pixels are shown as blue dots and the saturated pixels are shown as red diamonds. The PSL values for unsaturated pixels are used to construct a quadratic fit which is extrapolated to determine the actual signal. The correlation coefficient (R$^{2}$) for the fit is 0.98.}
\label{fig:saturation}
\end{figure}

\section{Results and discussion}

\begin{figure*}[Ht]
	\includegraphics[width=0.85\textwidth,natwidth=1000,natheight=400]{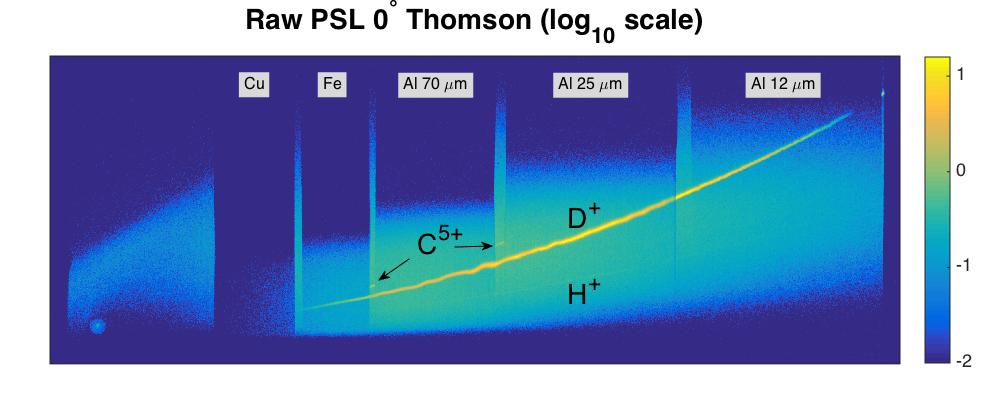}
	\caption{(color online). Raw PSL counts from a sample shot. The bright yellow quadratic track is the deuterium ion signal (labeled D$^{+}$). The extremely faint proton signal (labeled H$^{+}$) is below the deuterium track. The regions corresponding to the differential filters are labeled (Cu, Fe, etc.) at the top. There are gaps between each filter - faint C$^{5+}$ tracks are visible in the Fe/Al 70$\mu$m and the Al 70$\mu$m/Al 25$\mu$m gaps.}
	\label{fig:TP3_raw_spec}
\end{figure*}

Raw PSL counts (log$_{10}$ scale) from a sample shot (0$^{\circ}$ spectrometer) is shown in figure \ref{fig:TP3_raw_spec}. The bright parabolic line is the deuterium signal (labeled D$^{+}$); below it a very faint proton line (labeled H$^{+}$) is visible. The filters (Cu, Fe, Al 70 $\mu$m, etc.) used in each region are labeled at the top. Between each filter there is a gap of approximately 1mm; in the Fe/Al 70$\mu$m and Al 70$\mu$m/Al 25 $\mu$m gaps there is a faint O$^{+}$ track visible. The three other TPS recorded qualititatively similar results; strong deuterium ion signal and nearly extinguished contaminant signal.

\begin{figure}[thb]
	\includegraphics[width=0.49\textwidth,natwidth=1000,natheight=700]{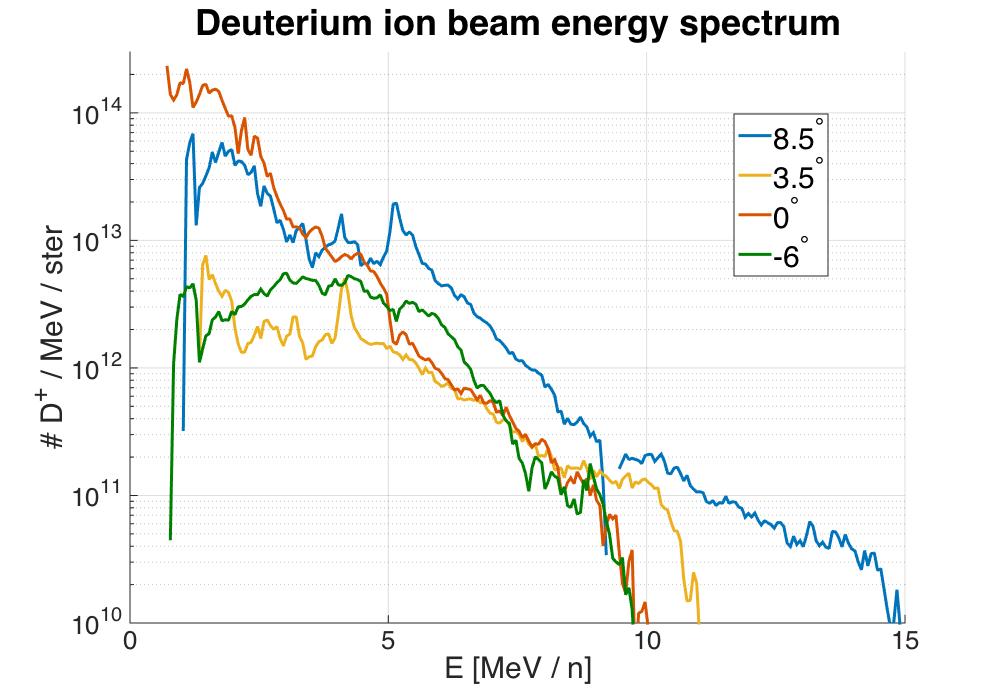}
	\caption{(color online). Energy spectra recorded by the TPS positioned at 8.5$^{\circ}$ (blue), 3.5$^{\circ}$ (yellow), 0$^{\circ}$ (red), and -6$^{\circ}$ (green).  The horizontal axis is given in MeV/nucelon.}
	\label{fig:d_spectrum}
\end{figure}

The absolute energy spectra from all 4 detectors are shown in Figure \ref{fig:d_spectrum} (same shot as Figure \ref{fig:TP3_raw_spec}). The maximum energy deuterium ion, recorded by the 8.5$^{\circ}$ TPS, was 14 MeV/nucleon. The deuterium ion-to-proton ratio for the spectrometer in Figure \ref{fig:TP3_raw_spec} is better than 0.99, consistent with the purity of the source heavy water. Across 5 shots (20 recorded spectra), a ratio of $\sim$0.9 or better is typical and the minimum observed ratio is $\sim$0.7. The differential filters blind the detectors to heavier ions but one can assume that these heavier ions are also minimized. This assumption is based on the relatively low signal observed in the filter gaps and the minimal proton signal. Protons should be preferentially accelerated over other species because they are the lightest ions of all candidates; their absence is strong evidence for minimization of other contaminant species.

Finally, we estimate the deuterium ion beam properties above 0.7 MeV/nucleon, the minimum energy incident on the detector. Inside our detector limit, the conversion efficiency is 4.3$\%$, found by assuming azimuthal symmetry and making a linear fit to the TPS over the polar angle. However, this is likely a significant underestimate due to the angular cut-off.  The full-beam conversion efficiency can be estimated by considering the typical divergence of ions reported in literature for similar conditions. For instance, Maksimchuk et al., using a similar heavy ice target configuration with a much smaller laser (6 J, 15 TW), observed a Gaussian angular distribution with a 10$^{\circ}$ HWHM.\cite{Maksimchuk:PRL2000} Using these results, we find a conversion efficiency of 9.4$\%$. On the other hand, N\"{u}rnberg et al.\cite{Nurnberg:RSI2009} observed a proton beam that was much broader than Maksimchuk. These results used the same plasma mirror configuration at Vulcan PW as ours suggesting that 9.4$\%$ efficiency could be an underestimate. We are planning to characterize the full energy-resolved angular distribution as well as repeatability in a future experiment.

\section{CONCLUSIONS}

We have demonstrated the ability to produce a nearly pure deuterium ion beam from cryogenic Au targets coated by a layer of heavy ice with a high-energy petawatt laser. \emph{Inside our detector limit} (0 - 8.5$^{\circ}$) we observe an ion beam with $>$0.99 deuterium-to-proton yield ratio, high peak energy (14 MeV/nucleon), and high conversion efficiency (4.3$\%$); a conservative estimate for the total conversion efficiency is 9.4$\%$. Further investigations will be done to fully characterize the deuterium ion beam in the future.

\section{ACKNOWLEDGEMENTS}

We are grateful for the dedicated work of laser operators, engineers, and technicians at Rutherford Appleton Laboratory who supported this experiment. A.G.K. and R.R.F. acknowledge funding from the United States Department of Energy through contract DE-FC02-04ER54789. J.T.M acknowldeges funding from the Quantum and Non-Equilibrium Processes Department of the Air Force Office of Scientific Research, under the management of Dr. Enrique Parra, Program Manager. We also acknowledge funding from EPSRC, through grants EP/J002550/1-Career Acceleration Fellowship held by S.K., EP/L002221/1 and 
EP/K022415/1. M.B. acknowledges co-financing by by the European Social Fund and the state budget of the Czech Republic (project numbers CZ.1.05/1.1.00/483/02.0061 and CZ.1.07/2.3.00/20.0279).

\end{document}